\documentclass[11pt,letterpaper]{article}
\usepackage{aaai17}

\usepackage{times}
\usepackage{latexsym}
\usepackage{booktabs}

\usepackage{graphicx}
\usepackage[usenames]{xcolor}

\newif\ifdraft
\drafttrue

\begin{document}

\title{QT2S: A System for Monitoring Road Traffic via Fine Grounding of Tweets} 

\author{Noora Al Emadi$^{\ddag}$, Sofiane Abbar$^{\ddag}$, Javier Borge-Holthoefer$^{\diamond}$, Francisco Guzm\'an$^{\ddag}$, Fabrizio Sebastiani$^{\heartsuit}$ \\
$^{\ddag}$Qatar Computing Research Institute,
Hamad bin Khalifa University, Qatar \\
$^{\diamond}$ Internet Interdisciplinary Institute, Universitat Oberta de Catalunya, Spain\\
$^{\heartsuit}$Istituto di Scienza e Tecnologie dell'Informazione, Consiglio Nazionale delle Ricerche, Italy \\
E-mail: \texttt{nalemadi@hbku.edu.qa}, \texttt{sabbar@hbku.edu.qa}, \texttt{jborgeh@uoc.edu},\\ \texttt{guzmanhe@gmail.com}, \texttt{fabrizio.sebastiani@isti.cnr.it}}

\maketitle

\begin{abstract}

Social media platforms provide continuous access to user generated content that enables real-time monitoring of user behavior and of events. The geographical dimension of such user behavior and events has recently caught a lot of attention in several domains: mobility, humanitarian, or infrastructural. While resolving the location of a user can be straightforward, depending on the affordances of their device and/or of the application they are using, in most cases, locating a user demands a larger effort, such as exploiting textual features. On Twitter for instance, only 2\% of all tweets are geo-referenced. In this paper, we present a system for zoomed-in grounding (below city level) for short messages (e.g., tweets). The system combines different natural language processing and machine learning techniques to increase the number of geo-grounded tweets, which is essential to many applications such as disaster response and real-time traffic monitoring.

\end{abstract}

% -----------------------------------------------------------------

\section{Introduction}

\noindent Social media platforms provide continuous access to information generated by users around the world. This enables real-time monitoring of user behavior \cite{abbar_chi2015}, events \cite{WengL11}, and urban dynamics \cite{abbar2016}. The geographical dimension of such user behavior and events has recently caught a lot of attention, and for different reasons: commercial, humanitarian, or infrastructural. In some cases, resolving the location of a user is straightforward, depending on the affordances of their device and/or of the application they are using (e.g. GPS-enabled activity). However, in most cases, locating a user demands a larger effort, such as exploiting textual content of web entries \cite{rae2012mining} or social media messages to estimate location \cite{han2014text}. 

Researchers have dealt with this problem with varying success. State-of-the-art techniques for dealing with free text \cite{rae2012mining} reach above $90\%$ precision at coarse-grained {\em grounding} of a user's activity (e.g. country, city levels), which suffices for many applications (say, to place ads online).  However, {\em zoomed-in} grounding (below city level) is essential if the purpose is to profile traffic conditions or to assess damage resulting from a natural disaster.

In this paper we present a system which instantiates this fine grounding. We do so by presenting a dedicated implementation, QT2S (Qatar Traffic Social Sensing), which illustrates its usefulness to locate traffic-related events (e.g., traffic jams) in Doha, the capital of Qatar. 
While some location based social networks such as Instagram present higher fractions of geo-coded posts (30\%) compared to Twitter (2\%) \cite{MejovaAH16}, it is important to notice that these services do not open their data to third parties.  
The idea then is to exploit the user generated content widely available in open social media platforms (Twitter) in order to enrich with more semantics traditional traffic monitoring systems based on physical counting sensors, which often lack the semantics of what is causing congestion to happen (accidents, traffic deviations, malfunctioning signals). Our work first aim is to fill the gap of below-city-level geo-location, and we do so in a challenging setting: First, the population of Qatar is linguistically heterogeneous and is composed by over 85\% of migrants (many of whom do not speak Arabic); second, there is no standard for transliteration from Arabic into English, which adds ambiguity (esp. when referring to local named entities) to the already challenging informal content of social media; third, the rapid growth of the city and the difficulties that migrants experience in dealing with Arabic names of public spaces, have led to a landmark-driven navigation, which results in many places being referred to by multiple expressions. 
%

% -----------------------------------------------------------------

\section{System Architecture}

\noindent In this section we present an overview of our system, first describing its  architecture and then proceeding with the description of each module. 

\begin{figure}[htp]
 \centering
 \includegraphics[width=0.98\linewidth]{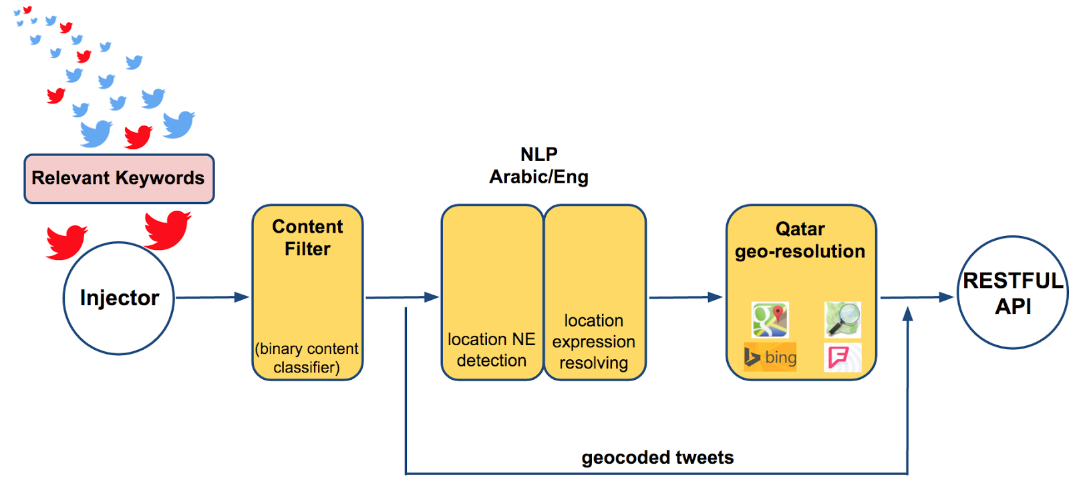}
 \caption{The data processing pipeline.} 
%  \fab{I believe that the figures should be displayed in greyscale, and not in color, according to NAACL rules.}
 \label{fig:architecture}
\end{figure}

The general view of the system is depicted in Figure \ref{fig:architecture}. The process starts by listening to the Twitter feed in order to catch relevant social posts using a handcrafted list of 70 keywords related to the context of the system (e.g., road traffic). %\fab{Sofiane or Javier, how many keywords does that list contain? It would be nice to indicate it.} 
Returned posts are then pushed through a three-steps pipeline in which: (i) we double-check the relevance of the post using a binary classifier (the \emph{Content Filter}), (ii) we extract location names mentioned in the posts, if any, and (iii) we geo-locate the identified locations to their accurate placement on the map. This process allows to filter undesirable posts, and augment the relevant ones with precise geo-location coordinates which are finally exposed for consumption via a RESTful API. %Below we provide details on each of the aforementioned modules. 

% -----------------------------------------------------------------

\subsection{Content filter}

The \textit{Content Filter} takes the stream of tweets as input and preserves only traffic-related tweets. It consists of two different modules: a keyword-based filter, and a binary classifier that further refines the selection. 
The keyword-based filter selects tweets that match one or more keywords that refer to traffic-related events (e.g., \emph{accident}, \emph{traffic}, \emph{jam}, etc.). However keyword-based filtering is a noisy criterion to select events related to traffic. Therefore, we further train a binary classifier to filter out tweets that are not real-time reports about traffic (e.g., tweets that mention \emph{jam} as a type of food, tweets that complain about the \emph{traffic} in general, etc.). To build the classifier, we use the Stanford maximum entropy (Max-Ent) classifier
\footnote{http://nlp.stanford.edu/software/classifier.html}, using with n-gram features. For its training, we manually annotated 1200 tweets. Each tweet is tokenized using a Twitter-specific tokenizer \cite{OConnorKA10}. Next, a rule-based local named entity simplifier is used to substitute mentions of local entities by their corresponding meta-categories (for example, it substitutes \emph{@moi\_qatar} -- which stands for the Qatar Ministry of Interior -- with the tag \emph{government\_entity}).  
  
In Table \ref{tab:cofi} we present a performance comparison between the keyword-based system and the system in which the results of the former are fed into the Max-Ent classifier. Results are calculated using 10-fold cross-validation on the training set. Initially, the keyword-based filter incorrectly classifies as traffic-related two-thirds of the tweets. Using the Max-Ent classifier we double the precision for the positive tag, thus avoiding to push many irrelevant tweets downstream. The overall accuracy increases from  0.337 to 0.772. 

\begin{table}[h]
\centering
\begin{tabular}{lcccc}
\toprule
& Prec & Rec & $F_1$ & Acc \\
\midrule
Keyword-based   & 0.337 & 1.000  & 0.505  & 0.337 \\
Maximum-entropy & 0.668 & 0.646  & 0.657  & 0.772 \\
\bottomrule
\end{tabular}
\caption{\label{tab:cofi}Positive label precision, recall, $F_1$ and overall accuracy for the two types of content filter.}
\end{table}

% -----------------------------------------------------------------

\subsection{Location Expression Recognition}

\noindent Qatar is a region where locations are rarely indicated by their street addresses. One of the main challenges in such a region is that locations and directions are thus given in relation to landmarks or distinctive points of interest. Therefore, to geocode an event (e.g., a traffic jam) described in a specific tweet, we  first need to identify all the possible locations involved. The \textit{Location Expression Extractor} is a module that identifies (or extracts) location expressions, i.e., natural language expressions that denote locations 
(e.g., \emph{at the Slope roundabout}, \emph{on Khalifa St}, etc.). Location expressions can be complex linguistic objects that involve several named entities of type Location, e.g., \emph{on} \textsf{location}$_A$ \textit{between} \textsf{location}$_B$ \textit{and} \textsf{location}$_C$. A key component of the Location Expression Extractor is the \textit{Location Named Entity Recognizer}, which identifies named entities of type Location  (e.g., \emph{the Sports roundabout}) or Landmark (e.g., \emph{the Museum of Islamic Art}). 

For our purposes, a location is any proper name in the Doha street system (e.g., \emph{Corniche}, \emph{TV Roundabout}, \emph{Khalifa Street}, \emph{Khalifa}); landmarks are different from locations, since  locations are only functional to the Doha street system, while landmarks have a different purpose (e.g., the MIA is primarily a museum, although its whereabouts may be used as a proxy of a specific location in the Doha street system, i.e., the portion of the Corniche -- the main Doha seafront road -- that is right in front of it). A key difference, for our purposes, is that a landmark can be mentioned in a tweet for a purpose other than indicating a place, as in \textit{The Traffic Department ruled that ...} or \textit{the MIA hosts a new exhibition}, while the same cannot be said of entities such as \textit{the Majlis Al Taawon Intersection}. Names of streets, roads, intersections, roundabouts, squares, are all location names; names of buildings, shops, parks, institutions, are all landmark names.

The Location Named Entity Extractor receives as input the set of tweets about traffic-related events in Doha, and returns the same tweets where named entities of type Location or of type Landmark have been marked as such. We generate a Location Named Entity Extractor via 
the Stanford CRF-based Named Entity Recognizer \cite{Finkel_20055}, using POS tags from the CMU-Ark Twitter POS tagger \cite{gimpel,OConnorKA10} using the tag-set from \cite{ritter_2011:EMNLP} and a location-specific gazetteer containing more than 700 locations and street names in Doha. The gazetteer is obtained by leveraging different sources, such as Foursquare locations and Open Street Map data.
For training we used 400 manually annotated tweets, containing 7.3K words and 1.3K location tags. For testing we used 170 tweets, containing 3K words and 560 location tags.  In Table \ref{tab:ner} we display the results for the Twitter-specific NER, which show how the performance of the recognizer improved dramatically from a previous version which was based on gazetteer matching only and used no CRF learning.

% -----------------------------------------------------------------

%\subsubsection{Performance evaluation}

\begin{table}[h]
\centering
\begin{tabular}{lcccc}
\toprule
&Prec&Rec& $F_1$ \\
\midrule
Gazeteer & 0.810 & 0.120 & 0.210 \\
CRF      & 0.896 & 0.674 & 0.769 \\
\bottomrule
\end{tabular}
\caption{\label{tab:ner} Word-level precision, recall and $F_{1}$ for Location NER.}
\end{table}

% -----------------------------------------------------------------

\begin{figure*}[ht]
 \centering
 \includegraphics[width=0.95\linewidth]{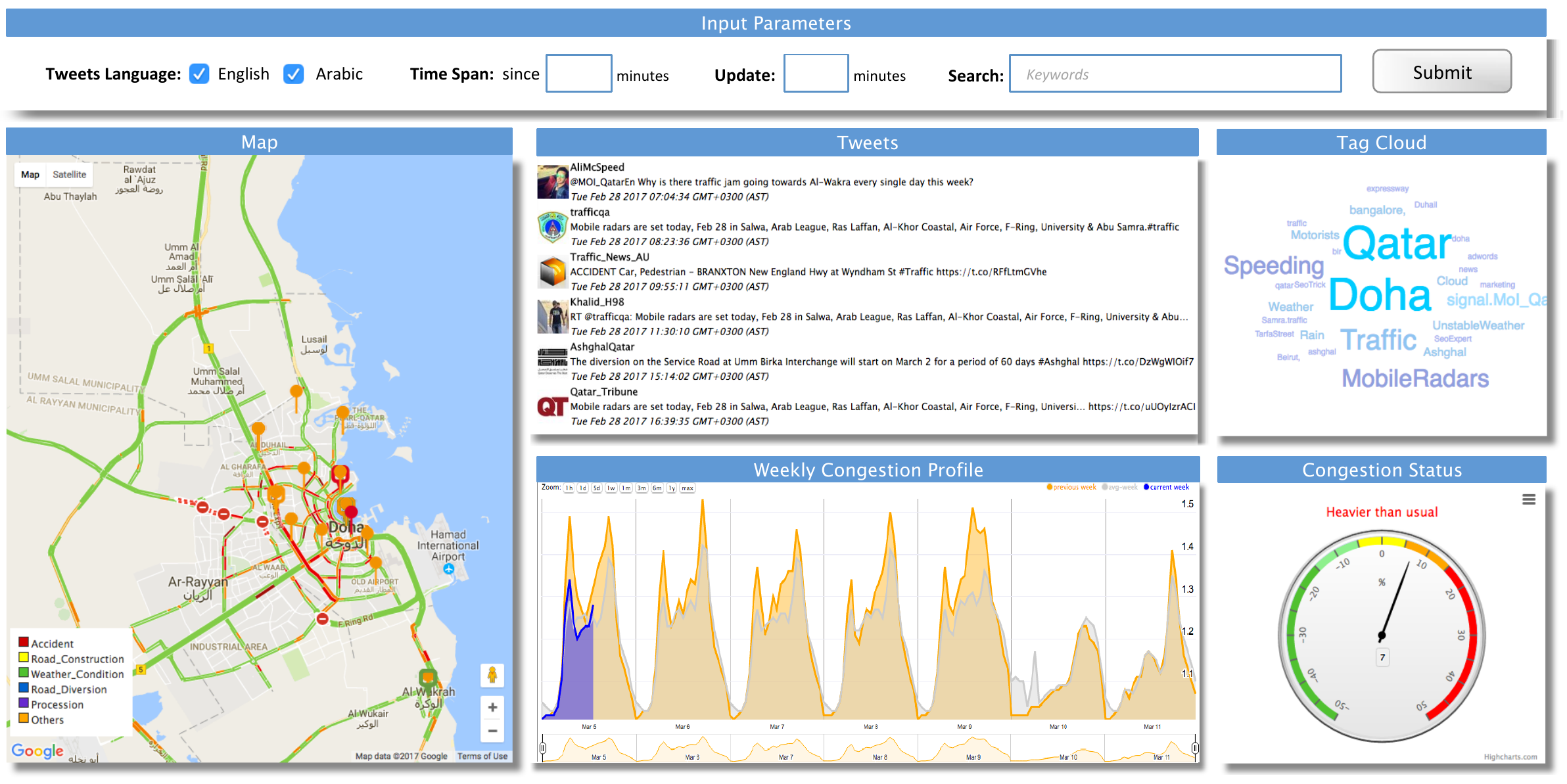}
 \caption{Snapshot of some of QT2S frontend widgets.} 
 \label{fig:dashboard}
\end{figure*}

\subsection{Resolving location expression onto the map}

Given a location entity identified by the previous module, we first check whether or not the location exists in the gazetteer by using approximate string matching. If the location is not found, the geocoding APIs of Google\footnote{http://bit.ly/1NZn2wa (accessed on November 15, 2015)}, Bing\footnote{https://msdn.microsoft.com/en-us/library/ff701711.aspx (accessed on November 15, 2015)} and Nominatim\footnote{http://wiki.openstreetmap.org/wiki/Nominatim (accessed on November 15, 2015)} are used to map the new location into a pair of latitude / longitude coordinates. We use multiple geocoding sources to increase the robustness of our application, since a single API might fail to retrieve the geographic coordinates of the identified location.

The result  geocoding is formatted as a JSON object containing the name of the location entity, its address, and the corresponding geocoding results from Bing, Google and Nominatim. The geocoding process is validated by comparing the results of the different services used. We first make sure that the location returned falls within Doha's bounding box. We then compute the pairwise distance between the different geographic coordinates to ensure their consistency. If the coordinates returned by the three services are far apart from each other, then the location is ignored.  

While it is hard to evaluate the accuracy of such a system, since it relies on many external services, we found that combining our NLP components with geocoding services allows geo-grounding many traffic-related tweets. 
E.g., from 2015/10/09 to 2015/11/12 the system processed $\approx$25M tweets matching one of the keywords we used to filter the Twitter stream. Among them, 3K only have made it to the map, with only 143 gps-geocoded tweets and 2857 tweets grounded by our system. 
That is, our NLP pipeline augmented the number of available tweets with precise locations 20x. Note that for gps-geocoded tweets, location is about the place where the tweet was issued from; whereas detected locations by QT2S are those referred-to in the text. Obviously, those two locations may not coincide. 

% -----------------------------------------------------------------

\subsection{Description of the RESTful API}

\noindent To facilitate the consumption of the relevant geo-located posts and make it possible to integrate them in a comprehensive way with other platforms, we have built a RESTful API. In the context of our system, this refers to using HTTP verbs (GET, POST, PUT) to retrieve relevant social posts stored by our back-end processing.
Our API exposes two endpoints:\ Recent and Search. The former provides an interface to request the latest posts identified by our system, and supports two parameters: Count (maximum number of posts to return) and Language (the language of posts to return.) 
The latter endpoint enables querying the posts for specific keywords and return only posts matching them. It supports three parameters: Query (list of keywords), Since (datetime of the oldest post to retrieve), From-To (two datetime parameters to express the time interval of interest). In the case of a road traffic application, one could request tweets about ``accidents'' that occurred in ``West Bay'' since the 10th of October. 

% -----------------------------------------------------------------

\section{Repurposing the architecture for different applications}

\noindent Our proposed platform is highly modular (see Figure 1). This guarantees that relatively simple changes can make the platform relevant to any application context where anchoring user messages to specific locations on a map is needed. For instance, the content filter can be oriented to mobility problems in a city: accident or congestion reporting, road blocking or construction sites, etc. With a suitable classifier, our platform can collect traffic and mobility tweets, and geo-locate them when possible. However, there are many other contexts in which precise location is needed. E.g., in natural disaster management it is well known that people who have witnessed catastrophic events (floods, typhoons, etc.) use social media as a means to create awareness, demand help or medical attention \cite{Imran13}. Quite often, these messages may contain critical information for relief forces, who may not have enough knowledge of the affected place and/or accurate information of the level of damage in buildings or roads. Often, the task to read, locate on a map and mark is crowdsourced to volunteers \cite{Imran:2014zl}; we foresee that, in such time-constrained situations, our proposed technology would represent a way to support the work of crowdsourcers, or to replace them when they are not available. 
%Likewise, the system may be oriented towards other applications: weather conditions, leisure, etc.

% -----------------------------------------------------------------

\section{Demonstrator script}

\noindent Our demo is about an instantiation of the proposed architecture to real-time sensing of the status of traffic in Doha (Qatar). Figure \ref{fig:dashboard} shows a screenshot of QT2S dashboard and highlight five relevant visualization widgets along with a section for parameter selection and tunning:
\begin{itemize}
\item The \textit{Input Parameters.} Users can decide the data they want to render on the map by setting the following parameters: i) language of the tweets (so far the system supports Arabic and English). ii) Time interval of tweets and traffic data. iii) Refreshing frequency of the widgets. vi) Topic and keywords of interest. The last parameter allows users to explore a subset of Tweets that match some keywords.

\item The \textit{Map Widget.} Displays a map of Doha with different markers.
Colors of the markers represent different categories of tweets as labeled by our incident detection classifier (i.e. Accident, Road Construction, Weather Condition, ... etc). 
Large markers are used for tweets with attached images, whereas small markers are used for text-only tweets.
\item The \textit{Tag Cloud Widget.} Shows hashtags mentioned in the tweet collection corresponding to the input parameters of users. 
The font size is correlated with the hashtag frequencies.
\item The \textit{Tweets Widget.} Lists the traffic-related tweets which are collected by our system and that satisfy the input parameters of the user.
\item The \textit{Weekly Congestion Profile Widget.} Displays three weekly time series representing the congestion levels observed in Doha. Gray color is used for the time series is for the typical week, aggregated over the last six months. Orange color is used for the time series of the previous week. The blue colored time series correspond to the current week. Having the three time series superposed in one widget allows a user to better get a sense of the current traffic overhead.  
\item The \textit{Congestion Status Widget.} Shows the current overall congestion status in the city compared to the typical traffic congestion observed at the same hour of the weekday in the past six months. This value is updated every hour. 
\end{itemize}

% -----------------------------------------------------------------

\bibliographystyle{aaai}
\small
\bibliography{qt2s}

\end{document}